\documentclass[twocolumn,showpacs,preprintnumbers,amsmath,amssymb,floatfix,superscriptaddress]{revtex4}

\usepackage{color} 
\usepackage{graphicx}
\usepackage{dcolumn}
\usepackage{bm}

\begin{document}

\title{Elastic parton scattering and non-statistical event-by-event mean-$p_{t}$ fluctuations in Au + Au collisions at RHIC
}

\author{Qing-Jun Liu}
\affiliation{Department of Mathematics and Physics, Beijing Institute of Petro-chemical Technology, Beijing 102617, P.R. China}
\affiliation{CCAST(World Lab.), P.O. Box 8730, Beijing 100080, P.R. China}

\author{Wei-Qin Zhao}
\affiliation{CCAST(World Lab.), P.O. Box 8730, Beijing 100080, P.R. China}
\affiliation{Institute of High Energy Physics, Chinese Academy of Sciences, P.O. Box 918(4), Beijing 100039, P.R. China}

\begin{abstract}

        Non-statistical event-by-event mean-$p_{t}$ fluctuations in Au + Au collisions at
$\sqrt{s_{NN}} = 130$ and $200 $ GeV are analyzed in \textrm{AMPT} with string-melting, and the results
are compared with STAR data. The analysis suggests that in-medium elastic parton scattering
may contribute greatly to the mean-$p_{t}$ fluctuations in relativistic heavy-ion collisions.
Furthermore, it is demonstrated that non-statistical event-by-event mean-$p_{t}$ fluctuations can be used to probe
the initial partonic dynamics in these collisions. The comparison shows that with an in-medium elastic parton
scattering cross section $\sigma_{p}=10$ mb, \textrm{AMPT} with string-melting can well
reproduce $\sqrt{s_{NN}} = 130$ GeV data on the centrality dependence of non-statistical event-by-event
mean-$p_{t}$ fluctuations. The comparison also shows that the fluctuation data for $\sqrt{s_{NN}} = 200$ GeV
Au + Au collisions can be well reproduced with $\sigma_{p}$ between  $6$ and $10$ mb.

\end{abstract}

\pacs{25. 75. -q, 25. 75. Gz}
\maketitle

\section{ Introduction}
In order to detect the production and to study the properties of Quark-Gluon-Plasma(QGP),
non-statistical event-by-event mean-$p_{t}$ fluctuations in relativistic heavy-ion collisions have been
measured\cite{ceres,wa98,na49,star_reid,phenixFpt,star_westfall} at CERN SPS and BNL RHIC, and
have been extensively studied
theoretically\cite{phipt,Stodolsky,Shuryak,Gbaym,sigdy,Mjt,Trainor,RKorus,QjLiu,GavinS,PajaresFM,Voloshin,WBroniowski,JYCheng}.
While experimental data show that a strongly interacting quark-gluon-plasma(sQGP) has been
produced in heavy-ion collisions at RHIC \cite{RHICoverviewTH,RHICoverviewEXP}, a common
consensus on the explanation for the data on non-statistical event-by-event mean-$p_{t}$ fluctuations
measured at RHIC has not been reached
yet\cite{star_reid,phenixFpt,star_westfall,QjLiu,GavinS,PajaresFM,Voloshin,WBroniowski}.
To be more specific, the observed non-statistical event-by-event mean-$p_{t}$
fluctuations at RHIC currently can be explained in various physical scenarios: quenched
jets/minijets production \cite{phenixFpt,QjLiu},
a large degree of thermalization\cite{GavinS}, percolation of
strings\cite{PajaresFM}, a build-up of radial collective
flow \cite{Voloshin} and formation of "lumped clusters" \cite{WBroniowski,WBroniowski2}.
Though jets/minijets contribute significantly to the
event-by-event mean-$p_{t}$ fluctuations\cite{WBroniowski} and jet-quenching reduces the
fluctuations\cite{phenixFpt}, as \textrm{HIJING} model\cite{WangHijing} predicts\cite{QjLiu},
\textrm{HIJING} model significantly underestimates
the STAR data\cite{star_reid,star_westfall} on the event-by-event mean-$p_{t}$ fluctuations.
This may imply sources of the mean-$p_{t}$ fluctuations
beyond jets/minijets production. Taking into account the fact that
\textrm{HIJING} model includes neither partonic nor hadronic cascade processes,
one may wonder if the \textrm{HIJING}-based \textrm{AMPT} model\cite{AMPTdefault,amptSum},
which includes parton cascade together with hadronic cascade processes, is
able to reproduce the STAR data on non-statistical event-by-event mean-$p_{t}$ fluctuations.
Additionally, how partonic and hadronic cascade processes
contribute to the non-statistical event-by-event mean-$p_{t}$ fluctuations is an
interesting question in heavy-ion collisions at RHIC energies\cite{QjLiu}. The
intention of this letter is to answer these questions.

\section{ The \textrm{AMPT} model and analysis methods}
The \textrm{AMPT} model is a hybrid model. It uses minijet partons
from hard processes and strings from soft processes in
\textrm{HIJING}
as the initial conditions for modeling heavy ion collisions at
ultra-relativistic energies.
Since a phase transition of hadronic matter to quark-gluon plasma would occur
\cite{amptBin,Kharzeev} in Au + Au collisions at
RHIC, we study partonic effects on event-by-event mean $p_{t}$ fluctuations using
the AMPT model which allows the melting of the initial excited strings into partons\cite{Linzk}.
Interactions among these partons are
described by the\textrm{\ ZPC} parton cascade model\cite{ZhangZPC}.
At present, the AMPT model includes only parton-parton elastic scatterings
with an in-medium cross section given by:
\begin{equation}
\sigma _{p}\approx \frac{9\pi \alpha _{s}^{2}}{2
\mu ^{2}},  \label{crscp}
\end{equation}
where the strong coupling constant $\alpha _{s}$ is taken to be $0.47$.
The effective screening mass $\mu $ is generated by medium effects and given as
an adjustable parameter thus it can be used in studying effects of
parton scattering cross sections in heavy-ion collisions.
The transition from the partonic matter to the hadronic matter is
achieved using a simple coalescence model\cite{amptSum}.
Following the formation of hadrons, hadronic scatterings are then
modeled by a relativistic transport (\textrm{ART}) model \cite{Li1995}.
This version of the \textrm{AMPT} model is able to reproduce both the
centrality and transverse momentum (below $2$ \textrm{ GeV}$/c$) dependence
of the elliptic flow \cite{Linzk} and pion interferometry \cite{LinHBT02}
measured in Au + Au collisions
at \textrm{RHIC} \cite{Ackermann,STARhbt01}. It has been also applied
for studying kaon interferometry\cite{lin}, charm flow\cite{ZhangChFlow}, $\phi$ meson
flow\cite{JHChen} and and the system size dependence of elliptic
low\cite{ChenKo}.

In this letter, $\Delta \sigma_{p_{t}:n}$\cite{star_reid} and
$\langle \Delta p_{t,i} \Delta p_{t,j} \rangle$\cite{star_westfall} are
used in our analysis of event-by-event
mean-$p_{t}$ fluctuations.
According to Ref. \cite{star_reid,star_westfall}, they are defined as:
\begin{eqnarray}
\Delta \sigma_{p_{t}:n}=\frac{\Delta\sigma_{p_{t}:n}^{2}}{2\sigma_{\hat{p}_{t}}},
\end{eqnarray}
\begin{eqnarray}
\langle \Delta p_{t,i} \Delta p_{t,j} \rangle=\frac{1}{\epsilon}
\sum_{k=1}^{ \epsilon}{\frac{C_{k}}{n_{k}(n_{k}-1)}},
\end{eqnarray}
where
\begin{eqnarray}
\Delta\sigma_{p_{t}:n}^{2}=\frac{1}{\epsilon}
\sum_{k=1}^{\epsilon}{n_{k}[\left\langle {p_t} \right\rangle _{k}-\hat{p}_{t}]^{2} -\sigma_{\hat{p}_{t}}^{2}},
\end{eqnarray}
\begin{eqnarray}
C_{k}=\sum\limits_{i=1}^{n_{k}}
{
\sum\limits_{j=1,i\ne j}^{n_{k}}
{
{\left(
{p_{t,i}-
\left \langle \left\langle {p_t}
\right\rangle\right\rangle }
\right)}~
{\left(
{p_{t,j}-
\left\langle\left\langle {p_t}
\right\rangle\right\rangle }
\right)}
}
},
\end{eqnarray}
and $\epsilon$ is the number of events,
$n_k$ is the number of particles in the $k^{th}$ event,
$\hat{p_{t}}$ and $\sigma_{\hat{p_{t}}}$ are the mean and variance of the inclusive
transverse momentum distribution.
$\left\langle {p_t} \right\rangle _k$ is
the average transverse momentum for the $k$th event defined as
\begin{eqnarray}
\left\langle {p_t} \right\rangle _k=\left(\sum\limits_{i=1}^{n_{k}}
{p_{t,i}}\right) /n_k,
\end{eqnarray}
where $p_{t,i}$ is the transverse momentum of the $i^{th}$ particle in that event.
$\left \langle \left \langle p_{t} \right \rangle \right \rangle$ denotes the mean of
the $\left \langle {p_t} \right\rangle$ distribution and is given by
\begin{eqnarray}
\left \langle \left \langle p_{t}\right \rangle \right \rangle=\left( {\sum\limits_{k=1}^{ \epsilon}}
{\left\langle {p_t} \right\rangle _{k}} \right)/\epsilon.
\end{eqnarray}
As pointed out in Ref.~\cite{star_reid,star_westfall,WBroniowski2}, both $\Delta \sigma_{p_{t}:n}$ and
$\langle \Delta p_{t,i} \Delta p_{t,j} \rangle$, together with other event-by-event mean-$p_t$ fluctuation measures,
such as $F_{p_t}$\cite{phenixFpt} and $\Phi_{p_t} $\cite{phipt},
are by definition determined by two-particle transverse momentum correlations.
When only statistical event-by-event mean-$p_{t}$ fluctuations exist, there are no two-particle transverse
momentum correlations, hence a null value would be obtained for $\langle \Delta p_{t,i} \Delta p_{t,j} \rangle$
and $\Delta \sigma_{p_{t}:n}$. For more details regarding the two measures, interested readers are
referred to Ref. \cite{star_reid,star_westfall,sigdy,Trainor}.
\section{Sources of non-statistical event-by-event mean-$p_{t}$ fluctuations}
\begin{table}[b]
\caption{ $\langle \Delta p_{t,i} \Delta p_{t,j} \rangle$(MeV/c)$^{2}$,
together with $\Delta \sigma_{p_{t}:n}$(MeV/c) as well as the total number of events($\epsilon$),
the mean ($\hat p_t$) and variance ($\sigma_{\hat p_t}$) of inclusive transverse momentum
distributions, the mean ($\bar{n}$) and variance ($\sigma_{n}$) of multiplicity distributions,
in-medium elastic parton scattering cross section($\sigma_{p}$)
for four classes of mid-central \textrm{AMPT} events($5 < b < 7 \it{fm}$)---Au + Au collisions
at $\sqrt{s_{NN}}~=~130~$ GeV. Except for $\sigma_{p}$ and $\epsilon$, these quantities are
calculated using charged hadrons with $|\eta|<1$ and $0.15 < p_{t}<2$ GeV/c. Errors are statistical only. \label{TableI}}
\begin{tabular}{|c|c|c|c|c|}
\hline
event class & I & II& III & IV \\
\hline
$\sigma_{p}$(mb) & $10$ & $10$& N/A & $3$ \\
\hline
$\epsilon$ & $50,686$ & $50,480$ & $45,000$ & $31,981$ \\
\hline
$\bar{n}$& 600 & 607 & 765 & 616 \\
\hline
$\sigma_n$ &90& 90 & 122 & 90 \\
\hline
$\hat p_t$(MeV/c) & 479 & 477 & 458 & 457 \\
\hline
$\sigma_{\hat p_t}$(MeV/c) & 285 & 276 & 263 & 275 \\
\hline
$\Delta \sigma_{p_{t}:n}$ & $70.3 \pm 1.3$ & $79.9 \pm 1.4$ & $29.7 \pm 1.1$ & $31.7 \pm 1.4$ \\
\hline
$\langle \Delta p_{t,i} \Delta p_{t,j} \rangle$ & $68.5 \pm 1.3 $& $73.9 \pm 1.3$ &$21.1 \pm 0.8$ & $29.0 \pm 1.2$\\
\hline
\end{tabular}
\end{table}
       Particle production in ultra-relativistic heavy-ion collisions involve the following stages:
the initial stage, where jets/minijets are produced followed by partonic interactions resulting in the
formation of a partonic medium even quark-gluon-plasma; then the partonic medium undergoes hadronization to
become hadrons; what follows is a hadronic stage when the hadrons may decay or interact with each
other experiencing hadronic cascade until freeze-out to produce final state particles.
With AMPT, the following three classes of \textrm{AMPT} events may be generated: class I--events
with both partonic interactions and hadronic cascade process, class II--events with partonic interactions but
without hadronic cascade process, class III--events with neither partonic interactions nor hadronic cascade
process. Events of class III are primarily \textrm{HIJING} events with jet production but without jet quenching.
Therefore, through analyzing non-statistical event-by-event mean-$p_{t}$ fluctuations in the
aforementioned three classes of AMPT events, one can study how the non-statistical event-by-event
mean-$p_{t}$ fluctuations are built up through aforementioned stages of particle production processes. In another
word, using \textrm{AMPT}, one can learn to what extent jets/minijets, partonic interactions
and hadronic cascade process contribute to the event-by-event mean-$p_{t}$ fluctuations.

        We have generated events of those three classes for Au + Au collisions
at $\sqrt{s_{NN}}$ = 130 GeV with impact parameter of $5<b< 7$ fm, and analyzed non-statistical
event-by-event mean-$p_{t}$ fluctuations in each of the three classes of events for charged hadrons.
The results are tabulated in Table. \ref{TableI}. From Table. \ref{TableI}, one can see,
first of all, that in events of class III, there exist noticeable non-statistical event-by-event
mean-$p_{t}$ fluctuations due to production of jets/minijets\cite{QjLiu}.
It is noted that the values of the {\it a} and {\it b} parameter for string fragmentation used
in AMPT\cite{amptSum} is different from those in the \textrm{HIJING} model. This is why $\bar{n}$ for events of
class III is greater than that from \textrm{HIJING} without jet-quenching.  Secondly, the magnitude
of $\Delta \sigma_{p_{t}:n}$ in events of class II is more than two times the magnitude of
$\Delta \sigma_{p_{t}:n}$ in events of class III, and so is the magnitude of
$\langle \Delta p_{t,i} \Delta p_{t,j} \rangle$. This indicates that in addition to the production
of jets/minijets of partons, elastic parton scatterings add significantly to
the non-statistical event-by-event mean-$p_{t}$ fluctuations. Hence elastic parton scatterings may well be
among others a major source of the non-statistical event-by-event mean-$p_{t}$ fluctuations. Thirdly,
comparing the magnitudes of $\Delta \sigma_{p_{t}:n}$ and $\langle \Delta p_{t,i} \Delta p_{t,j} \rangle$ for
event class-I with those for class-II, one notices that due to hadronic cascade processes, the non-statistical
event-by-event mean-$p_{t}$ fluctuations built up in the initial stage,
which is characterized by the production of jets/minijets and parton scatterings,
decrease about 12\% and 7\% for $\Delta \sigma_{p_{t}:n}$ and
$\langle \Delta p_{t,i} \Delta p_{t,j} \rangle$, respectively. This reduction of the
mean-$p_{t}$ fluctuations is an indication that the late stage hadronic cascade processes that follow the
elastic parton scatterings dilute the correlations established in the initial stage. Therefore, one may
conclude according to Table. \ref{TableI} that in heavy-ion collisions at RHIC, non-statistical
event-by-event mean-$p_{t}$ fluctuations are generated in the initial stage; in addition to the production
of jets/minijets of partons, elastic parton scatterings also contribute a large portion of the non-statistical
event-by-event mean-$p_{t}$ fluctuations.

\section{Non-statistical event-by-event mean-$p_{t}$ fluctuations and $\sigma_{p}$ }
\begin{figure}[t]
\includegraphics[width=3.35in, height=3.0in]{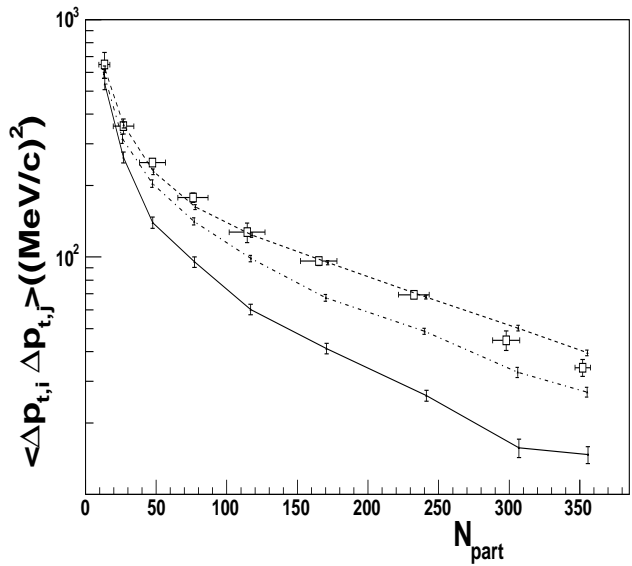}
\caption{\label{fig:fig1}
(Color online)$\langle \Delta p_{t,i} \Delta p_{t,j} \rangle$
in \textrm{AMPT} for $\sigma_{p}=10$ \textrm{mb}(dashed line)
, $\sigma_{p}=6$ \textrm{mb}(dash-dotted line) and $\sigma_{p}=3$ \textrm{mb}(solid line) as a function
of centrality for $\sqrt{s_{NN}}=130$ GeV Au + Au collisions compared with STAR
data(hollowed squares) as reported in Ref. \cite{star_westfall}.}
\end{figure}

   Taking place in the initial stage of ultra-relativistic heavy-ion collisions,
in-medium partonic interactions help drive the colliding system toward the formation of QGP.
In addition, considering the possible modification of partonic interactions in the partonic medium,
the in-medium parton scattering cross section incorporates properties of the medium and thus represents
an important aspect of partonic dynamics. To study the correlation between in-medium elastic parton
scattering cross section and non-statistical event-by-event mean-$p_{t}$ fluctuations,
in Table. \ref{TableI} we have tabulated our results for AMPT events with
in-medium elastic parton scattering cross sections $\sigma_{p} = 10$ \textrm{mb},
 $\sigma_{p} = 3$ \textrm{mb}. Comparing the magnitude of
$\Delta \sigma_{p_t}$ as well as $\langle \Delta p_{t,i} \Delta p_{t,j} \rangle$ calculated
using $\sigma_{p} = 10$ \textrm{mb} with that using $\sigma_{p} = 3$ \textrm{mb} as tabulated in
Table. \ref{TableI}, one may come to the
conclusion that non-statistical event-by-event mean-$p_{t}$ fluctuations increase significantly with the increase of
$\sigma_{p}$. This clearly indicates that
the event-by-event mean-$p_{t}$ fluctuations as measured with $\Delta \sigma_{p_t}$ or
$\langle \Delta p_{t,i} \Delta p_{t,j} \rangle$, are
rather sensitive to the in-medium elastic parton scattering cross section, and therefore are
promising probes of the partonic dynamics in the initial stage of ultra-relativistic heavy ion collisions.
\begin{figure}[t]
\includegraphics[width=3.35in, height=3.0in]{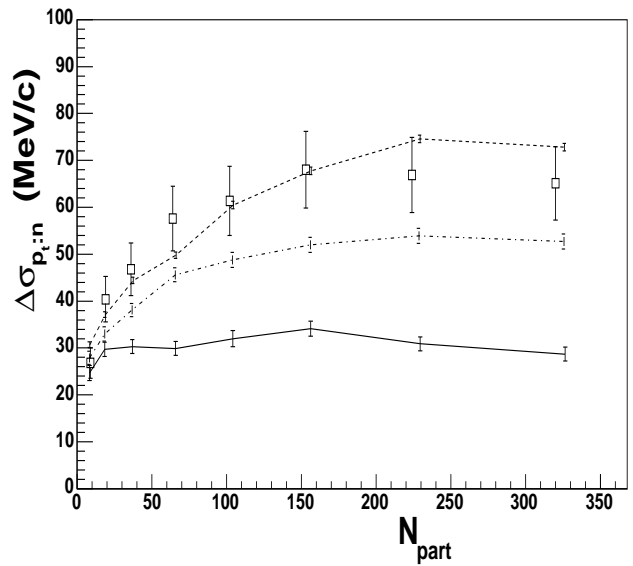}
\caption{\label{fig:fig2} (Color online)$\Delta \sigma_{p_{t}:n}$ as a function of centrality for Au + Au collisions
at $\sqrt{s_{NN}}=130$ GeV. The dashed line, dash-dotted line and solid line with error bars display \textrm{AMPT} results for
$\sigma_{p}=10$ \textrm{mb},$\sigma_{p}=6$ \textrm{mb} and $\sigma_{p}=3$ \textrm{mb}.
STAR data(hollowed squares) are as reported in Ref.\cite{star_reid}, being obtained via extrapolating
to 100\% tracking efficiency with the error bars representing $\pm$ 12\% uncertainties.}
\end{figure}

\begin{figure}[h]
\includegraphics[width=3.35in,height=3.0in]{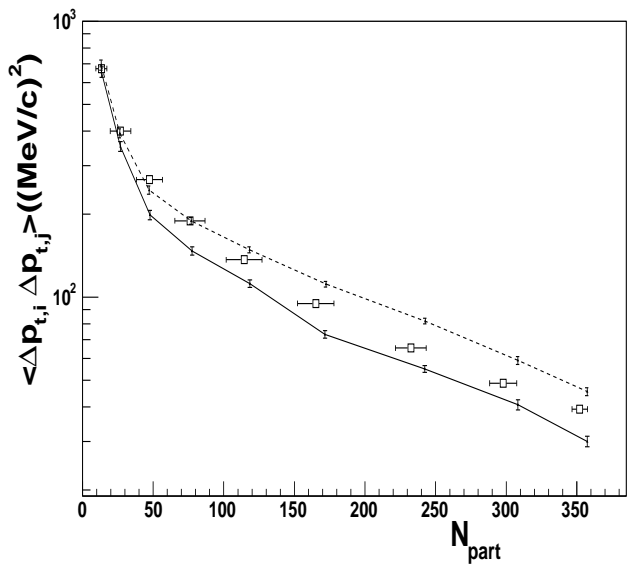}
\caption{\label{fig:fig3}
(Color online)$\langle \Delta p_{t,i} \Delta p_{t,j} \rangle$
in \textrm{AMPT} for both $\sigma_{p}=10$ \textrm{mb}(dashed line)
and $\sigma_{p}=6$ \textrm{mb}(solid line) as a function
of centrality for $\sqrt{s_{NN}}=200$ GeV Au + Au collisions compared
with STAR data (hollowed squares) as reported in Ref. \cite{star_westfall}.
}
\end{figure}

   To estimate the in-medium elastic parton scattering cross section $\sigma_{p}$ and to test
the \textrm{AMPT} model with string melting, we studied the centrality dependence of non-statistical
event-by-event mean-${p_{t}}$ fluctuations in Au + Au collisions at $\sqrt{s_{NN}}=130$ and $200$ GeV
based on \textrm{AMPT}. The results are compared with STAR data as is shown in Fig.~\ref{fig:fig1},
Fig.~\ref{fig:fig2} and Fig.~\ref{fig:fig3}.
In getting the \textrm{AMPT} results shown in Fig.~\ref{fig:fig1}, Fig.~\ref{fig:fig2} and Fig.~\ref{fig:fig3},
charged hadrons with transverse momentum 0.15 $\le p_{t} \le$ 2.0 GeV/c and pseudo-rapidity $|\eta| < $1.0
are used, the same cuts employed in obtaining the STAR data shown in these figures\cite{star_reid,star_westfall}. In calculating \textrm{AMPT} results for $\left \langle \Delta p_{t,i} \Delta p_{t,j} \right \rangle $ shown in Fig.~\ref{fig:fig1} and Fig.~\ref{fig:fig3}, the variation of $\left\langle {\left\langle {p_t} \right\rangle } \right\rangle$
within a given centrality bin is treated in the same way as Ref.~\cite{star_westfall} adopted to get the STAR data: $\left\langle {\left\langle {p_t} \right\rangle } \right\rangle$ is calculated as a function of $N_{ch}$, the multiplicity used to define the centrality bin. Then this dependence is fitted and the fit is used in Eq.~5 on an event-by-event basis as a function of $N_{ch}$.
Statistical errors for AMPT results are shown in Fig.~\ref{fig:fig1} through Fig.~\ref{fig:fig3}. 

        Both Fig.~\ref{fig:fig1} and Fig.~\ref{fig:fig2} demonstrate that with an elastic parton scattering cross
section $\sigma_{p} = 10$ mb the \textrm{AMPT} model with string-melting can reproduce STAR data on the
centrality dependence of the non-statistical event-by-event mean-$p_{t}$ fluctuations. This conclusion is
consistent with the one drawn through interferometry study of Au + Au collisions at the same
energy\cite{LinHBT02,lin}.
Fig.~~\ref{fig:fig1} and Fig.~\ref{fig:fig2} also reveal that for $\sigma_{p} = 3$ mb the \textrm{AMPT} model significantly underestimates the data
for non-peripheral collisions. For the most peripheral collisions, Fig.~\ref{fig:fig1} through
Fig.~\ref{fig:fig3} display good agreement between the data and the \textrm{AMPT} model
calculations with $\sigma_{p} = 6$ mb and $10$ mb. That
weak dependence on elastic parton scattering cross section $\sigma_{p}$ exists because in these peripheral
collisions elastic parton scatterings contribute trivially to the mean-$p_{t}$ fluctuations due to a very low parton density in these collisions.
From Fig.~\ref{fig:fig3}, one may infer that for Au + Au collisions
at $\sqrt{s_{NN}}=200$ GeV, the STAR data can also be well reproduced by the \textrm{AMPT} model and
support an estimate of $\sigma_{p}$ to be between $6$ and $10$ mb.
\section{Discussion}
 
     Jet/minijet partons may lose energy when passing through partonic medium. This parton energy loss called
jet-quenching usually consists of two parts: collisional energy loss and radiational energy loss. The parton
energy loss is modeled in the \textrm{AMPT}, as pointed out in Ref. \cite{amptSum}, via elastic two-body parton scatterings. However in this treatment, the radiational energy loss is ignored.
Looking at the AMPT results listed in Table. \ref{TableI}, by comparing $\langle \Delta p_{t,i} \Delta p_{t,j} \rangle$ and $\Delta \sigma_{p_{t}:n}$ from events of class II with elastic parton scattering to those of class III without elastic parton scattering,
we find that the elastic two-body parton scattering, while adding to
collisional parton energy loss, gives an additional event-by-event mean-$p_{t}$ fluctuations in mid-central collisions.
In the following a possible explanation is given for this result. First, based on the AMPT model, parton cascade process due to elastic parton
scattering is proposed to be a mechanism\cite{GLM,TEMS} for producing a Mach-like cone structure, which has recently been
observed in Au + Au collisions at RHIC\cite{MachE} and has been studied in several physical
scenarios\cite{GLM,TEMS,Stork,Casalder,vitev,Koch,Armesto,Ruppert,TRen,Chaud,Sat}. Secondly, the Mach-like cone structure
consists of clusters of final state charged particles and each of the cluster is consistent with conic emission
of particles from jets, in this sense it is jet-like and its formation represents an increase of the number of jet-like
clusters. Thirdly, in a recent study based AMPT, the number of correlated final state charged particles within the
clusters of the Mach-like cone structure is reported to increase with $\sigma_{p}$ changing from $3$ mb to $10$ mb\cite{TEMS}.
Furthermore, stronger parton cascade due to greater $\sigma_{p}$ also increases the number of correlated hadrons on the near-side of an energetic particle as one may infer according to Ref. \cite{TEMS}.  
To be brief, an energetic parton through successive elastic collisions with surrounding medium may couple many partons together therefore more partons are 
correlated with the increase of parton scattering cross section.
Besides, according to Ref. \cite{WBroniowski,WBroniowski2}, the event-by-event mean-$p_{t}$ fluctuations would increase with
both the number of clusters and the number of charged particles in the clusters.
Since a larger $\sigma_{p}$ increases both the number of jet-like clusters and the number of charged particles in the
jet-like clusters, it is natural that the event-by-event mean-$p_{t}$ fluctuations becomes greater
with $\sigma_{p}$ changing from $3$ mb to $10$ mb as seen in Table. \ref{TableI} and in
Fig.~\ref{fig:fig1} through Fig.~\ref{fig:fig3}.

     Now we turn to the other energy loss mechanism, the radiational energy loss. A study by PHENIX Collaboration at RHIC
and a study based on \textrm{HIJING} support the idea that parton energy loss tends to decrease the mean-$p_{t}$
fluctuations at RHIC\cite{phenixFpt,QjLiu}. In the HIJING model, the jet-quenching mechanism includes only radiational
parton energy loss. Therefore it seems that the radiational parton energy loss causes the decrease of the mean-$p_{t}$
fluctuations. Nevertheless, it would be worthwhile to investigate and compare the two mechanisms of the parton energy loss,
namely, the collisional energy loss and the radiational energy loss, and their implementation in both AMPT and HIJING more
thoroughly before making final conclusions about the effect of parton energy loss on the mean-$p_{t}$ fluctuations in
heavy-ion collisions at RHIC.

\section{Summary}

      Based on \textrm{AMPT} model with string-melting an analysis of non-statistical event-by-event
mean-$p_{t}$ fluctuations in Au + Au collisions at $\sqrt{s_{NN}}=130$ and $200$ GeV is presented. The analysis
suggests that in-medium elastic parton scatterings can contribute significantly to the mean-$p_{t}$ fluctuations and the
mean-$p_{t}$ fluctuations can be used as good probes to study the initial partonic dynamics in these collisions. The \textrm{AMPT} results are compared with STAR data. This comparison shows that using an in-medium
elastic parton scattering cross section $\sigma_{p}=10$ mb, predictions of the \textrm{AMPT} model are in good
agreement with the $\sqrt{s_{NN}}=130$ data on the centrality dependence of non-statistical event-by-event mean-$p_{t}$ fluctuations. The
comparison also shows that to reproduce the mean-$p_{t}$ fluctuation data at $\sqrt{s_{NN}}= 200$ GeV,
$\sigma_{p}$ is approximately between 6 and 10 mb.
\section*{Acknowledgments}
        We acknowledge Professor C.M. Ko, Z.W. Lin, B. Zhang, B.A. Li for using their \textrm{AMPT} code. We appreciate
helpful discussions with Professors C.M. Ko and T.A. Trainor. This work is partly supported by
National Natural Science Foundation of China(Wei-Qin Zhao); SRF for ROCS, SEM; Beijing Institute of Petro-chemical Technology and Supercomputing Center, CNIC, CAS.

\end{document}